\documentstyle[12pt]{article}

\begin{document}

\setcounter{page}{0}

\begin{titlepage}

\title{ Quantum Critical Point in the Spin Glass-Kondo Transition
       in Heavy Fermion Systems}

\author{Alba Theumann\\
    Instituto de F\'{\i}sica  - UFRGS\\
      Av. Bento Gon\c{c}alves 9500\\
      91501-970 Porto Alegre, RS, Brazil\\[4mm]
    B. Coqblin\\
    Laboratoire de Physique des Solides\\
    Universit\'e Paris - Sud\\
      91405 Orsay Cedex, France}

\date{}

\maketitle

\thispagestyle{empty}

\newpage

\begin{abstract}
The Kondo-Spin Glass competition is studied in a theoretical model
  of a Kondo lattice with an intra-site Kondo-type exchange
interaction treated within the mean-field approximation, an
inter-site quantum Ising exchange interaction with random
couplings among localized spins and an additional transverse field
$ \Gamma$ in the x-direction, which represents a simple quantum
mechanism of spin flipping, in order to have a better description
of the spin-glass state and in particular of the Quantum Critical
Point(QCP). Taking here a parametrization $ \Gamma= \alpha
J_{K}^{2}$ (where $ J_{K}$ is the antiferromagnetic Kondo
coupling), we obtain two second order transition lines from the
spin-glass state to the paramagnetic one and then to the
Kondo-state. For a reasonable set of the different
parameters, the two second order transition lines do not
intersect and end in two distinct QCP. The existence of QCP in the
Spin Glass-Kondo competition allows to give a better description
of the phase diagrams of some Cerium and Uranium disordered
alloys.

\normalsize

\end{abstract}
\vspace*{0.5cm}

\end{titlepage}

%\end{document}
%\renewcommand{\baselinestretch}{1.5}

\setcounter{page}{1}

\section*{1.\ Introduction}

    It is well known that there exists a strong competition
 between the Kondo effect on each site  of a Kondo lattice
 and  the magnetic ordering arising from the
RKKY interaction between magnetic atoms in heavy fermion systems.
  The Doniach diagram\cite{1} gives a good
description of this competition: the Neel temperature $ T_{N}$ is
firstly increasing with an increasing of the antiferromagnetic s-f
exchange interaction constant $J_{K} (> 0)$, then it is passing
through a maximum and finally it tends to zero at the "quantum
critical point" (QCP), with a second order transition at zero
temperature. Such a decrease of $ T_{N}$ down to the QCP has been
observed in many Cerium compounds, such as$ CeAl_{2}$\cite{2} or
$CeRh_{2}Si_{2}$\cite{3}, under pressure. We also know that the
Neel temperature starts from zero at a given pressure and incrases
rapidly with pressure in Ytterbium compounds, such as
$YbCu_{2}Si_{2}$\cite{4} or $YbNi_{2}Ge_{2}$\cite{5}, in good
agreement with the Doniach diagram.
 Above the QCP, there exists a very strong heavy fermion
character, but several possible behaviours, i.e. the classical
Fermi liquid one with eventually
a reduced Kondo temperature\cite{6,7} or different Non-Fermi-Liquid (NFL)
ones, have been observed in Cerium or Ytterbium
compounds\cite{8,9}.\\
    On the other hand, the disorder can yield a Spin Glass (SG) phase
in addition to the Kondo (mainly NFL) behaviour at low
temperatures around the QCP in disordered Cerium or Uranium
alloys. This is the case of the magnetic phase diagram of
$CeNi_{1-x}Cu_{x}$ alloys that has been extensively
studied\cite{10,11}, while the phase diagram of $
Ce_{2}Au_{1-x}Co_{x}Si_{3}$ alloys presents the sequence of
SG-AF-Kondo phases at low temperatures with increasing Cobalt
concentration\cite{12}. The three phases, AF,SG, and NFL have been
obtained at low temperatures for different concentrations in $
UCu_{5-x}Pd_{x}$ \cite{13} or in $ U_{1-x}La_{x}Pd_{2}Al_{3}$
\cite{14} disordered alloys. Thus, a SG-Kondo transition has been
observed experimentally with increasing  concentration in
disordered alloys around the QCP.

    The purpose of our work here is to present a theoretical model
that describes the QCP for the  spin glass-Kondo transition in Kondo
lattices.The SG-Kondo transition was theoretically studied in
a previous publication\cite{15} and also in the presence of
ferromagnetic ordering\cite{16} or antiferromagnetic
ordering\cite{17}, but the QCP was not described
 because we lacked a quantum mechanism that favored spin flipping.
We present here an improvement of the previous model in
order to obtain a good description of the QCP.

More precisely, in previous publications\cite{15,16,17}
the resultant RKKY interaction was introduced by means of
random, infinite range couplings
among the $S^{z}$ components of the localized spins as in the
Sherrington-Kirkpatrick (SK) model\cite{18} and by neglecting the
spin flip coupling of the transverse components.
By using functional integral techniques with a static Ansatz in
a replica symmetric theory, we obtained a magnetic phase diagram
in the $J_{K} vs T$ plane, that showed the three different phases:
paramagnetic,spin glass and Kondo.
In spite of its complication,
the model failed to describe a second order QCP at $T=0$, because,
by disregarding the spin flipping part of the Heisenberg Hamiltonian,
we suppressed the tunneling mechanism, and  magnetic ordering
occurs only along the z-axis.
 In order to introduce a spin flipping mechanism and to avoid
the intricacies
of the random Heisenberg model\cite{19},
in the present paper the Heisenberg-like coupling among
the three spin components induced by the RKKY interaction
is mimicked by a quantum Ising spin glass in a transverse field.
It consists in
an effective random interaction among the z-components,
as we considered in\cite{15},
while the spin flipping part is simulated by a uniform transverse
field in the x-direction.

The infinite range quantum Ising spin glass in a transverse field  $ \Gamma$
is one of the simplest models that presents a quantum critical point  and
it is equivalent to the model for a spin glass of quantum
rotors\cite{20,21}.
The dynamical properties at zero temperature are known and the existence
of the QCP is well established\cite{22}, while the phase diagram in
the $T vs \Gamma $ plane
 has been  studied by using the Trotter-Suzuki technique\cite{23},
and more recently\cite{24} by the use of two fermionic
representations of the spin operators in terms of Grassmann
fields\cite{25} that are more suitable to our purposes.
 As we did  previously in reference\cite{15}, the Kondo effect
 will be studied in
an approximation that is basically equivalent to the mean field
decoupling scheme \cite{6,26}. The static Ansatz in the study of
the spin glass transition is an approximation, similar to mean
field theory, that is appropriate to describe the phase diagram.
This is justified because in the model of M-components quantum
rotors, that is exactly soluble for infinite M, the critical line
is given by the singularity of the zero-frequency
mode\cite{20}.The same occurs with the quantum Ising spin glass in
a transverse field\cite{22}, where the singularity of the zero
frequency mode determines the critical point. That is the reason
why the static Ansatz, that describes only the zero frequency
mode, describes the critical line in the phase diagram, although
it would not give correctly the time dependence of the order
parameter.

 A related Hamiltonian has been considered in ref.\cite{27}
to describe NFL behaviour and a QCP in some heavy fermion
compounds, although there are essential differences between this
work and ours in the present paper and in ref.\cite{15}. In
ref.\cite{27} it is considered a system of isolated Kondo
impurities, each one with a separate electron reservoir, and
represented by the spin glass model of M-components quantum rotors
in the limit of large M, when the problem is exactly soluble. The
Kondo coupling provides the quantum mechanism and the transition
line in the phase diagram is determined by the singularity of the
zero frequency mode, displaying a QCP at zero temperature. The
Kondo effect is described there\cite{27} by isolated impurities
and displays a
continuous transition among different scaling regimes, in place of
the sharp second order transition in Kondo lattices\cite{1,6,26}

This paper is organized as follows: in Sect.2 we introduce the
model, in Sect.3 we discuss relevant results and we reserve Sect.4
for discussions and conclusion. We refer the reader to
ref.\cite{15} for details in the mathematical calculations.

\section*{2.\ The model }

We consider a Kondo lattice system with localized spins
$\vec{S_{i}}$ at sites $i=1 \cdots N$, coupled to the electrons of
the conduction band via a s-f exchange interaction. It is
necessary to introduce explicitly the resultant RKKY interaction
by means of a random, infinite range coupling among localized
spins like in the Sherrington Kirkpatrick (SK) model for a spin
glass\cite{18}. To describe the Kondo effect in a mean-field-like
theory it is sufficient to keep only the spin-flip terms
\cite{6,15,26} in the exchange Hamiltonian, while the spin glass
interaction is represented by the quantum Ising Hamiltonian and the
 transverse field $\Gamma$ in the x-direction
\cite{24}.The transverse field $\Gamma $ represents a simple
quantum mechanism of spin flipping and mimics the  more
complicated transverse part of the Heisenberg
Hamiltonian\cite{19}.

The Hamiltonian of the model is:

\begin{eqnarray}
{\cal H}=
{\cal H}_{K} + {\cal H}_{SG}
\label{2.1}
\end{eqnarray}

\begin{eqnarray}
{\cal H}_{K}= \displaystyle \sum_{k,s}\epsilon_{k}n_{ks}
+\epsilon_{0} \displaystyle
\sum_{i,s}n_{i}^{f}+J_{K}\displaystyle
\sum_{i}[S_{fi}^{+}s_{di}^{-}+S_{fi}^{-}s_{di}^{+}]
\label{2.2}
\end{eqnarray}

\begin{eqnarray}
{\cal H}_{SG}=- \displaystyle \sum_{i,j}J_{ij}S_{fi}^{z}S_{fj}^{z}
-\displaystyle 2\Gamma\sum_{i}S_{fi}^{x}
\label{2.3}
\end{eqnarray}
\noindent where $J_{K} > 0 , S^{\pm}= S^{x} \pm iS^{y}$ and

\begin{eqnarray}
{\vec S}_{fi}=
\displaystyle \sum_{s,s^{\prime}}f_{is}^{\dagger}
{\vec \sigma}_{s,s^{\prime}}f_{is^{\prime}}
\nonumber\\
{\vec s}_{di}= \displaystyle \sum_{s,s^{\prime}}d_{is}^{\dagger}
{\vec \sigma}_{s,s^{\prime}}d_{is^{\prime}}
\label{2.4}
\end{eqnarray}
$f_{is}^{\dagger},f_{is}(d_{is}^{\dagger},d_{i})$ are
creation and destruction fermion operators for electrons with
$s=\uparrow $ or $\downarrow$ in the localized (conduction) band.
We indicate by $ \vec{\underline \sigma} $ the Pauli matrices and
we have $n_{k\sigma}=d_{k\sigma}^{\dagger}d_{k\sigma}$ the
conduction electrons occupation number. The energies
$\epsilon_{0}(\epsilon_{k})$ are referred to the chemical
potentials $\mu_{f}(\mu_{d})$, respectively.

The coupling $J_{ij}$ in Eq. (\ref{2.3}) is an independent random
variable\cite{18} with gaussian distribution of zero mean and
variance $< J_{ij}^{2} > = 8J^{2}/N $ .\\
Functional integration techniques have proved to be a suitable
approach to describe phase transitions in disordered quantum
mechanical many-particle systems  \cite{25}. The static approximation
within this formulation consists in neglecting time fluctuations of
the order parameter, and when it is combined with the neglect of space
fluctuations it leads to the usual Hartree-Fock, mean field like
approximation. When dealing with the Hamiltonian in
Eq. (\ref{2.1})-Eq.(\ref{2.3}), we notice that in the limiting case
$J_{K}=0$ we obtain a  quantum Ising spin glass in a transverse field
that has been studied with the static approximation in  \cite{24},
while for $J=0$
we recover the mean field approximation that has been used
successfully to describe the Kondo lattice \cite{6,8,26}.
We follow closely the formalism of Ref. (\cite{15})
to write a Lagrangian formulation\cite{25} in terms of
anticommuting, complex
Grassmann variables $ \varphi_{is}(\omega) , \psi_{is}(\omega) $
associated to the conduction and localized electron fields, respectively,
together with the spinors
\begin{eqnarray}
\underline{\varphi}_{i}(\omega)=\left(\begin{array}{c}
\varphi_{i\uparrow}(\omega)\\
 \varphi_{i \downarrow}(\omega)\end{array}\right)
\hskip 2cm
\underline{\psi}_{i}(\omega)=\left(\begin{array}{c}
\psi_{i\uparrow}(\omega)\\
 \psi_{i \downarrow}(\omega)\end{array}\right)
\label{2.5}
\end{eqnarray}
where $ \omega = (2n+1)\pi $ are the Matsubara frequencies.
The partition function is now expressed
\begin{eqnarray}
Z=\displaystyle \int D(\varphi^{\dag}\varphi)D(\psi^{\dag}\psi)e^{A}
\label{2.6}
\end{eqnarray}
We now follow the same steps as
in Ref. (\cite{15,16,17})
and in the static, mean field like
approximation the action A may be written
\begin{eqnarray}
A=A_{0}+A_{K}+A_{SG}
\label{2.7}
\end{eqnarray}
with $ A_{0} $ being the action for non-interacting electrons
in a transverse magnetic field:
\begin{eqnarray}
A_{0}=\displaystyle\sum_{\omega}\displaystyle
\sum_{i,j}\underline{\psi_{i}^{\dagger}}
(\omega)[(i\omega-\beta\epsilon_{0}
+\beta\Gamma{\underline\sigma_{x}}]\delta_{ij}
\underline{\psi_{i}}(\omega)
+ \underline{\varphi_{i}^{\dagger}}(i\omega\delta_{ij}-\beta t_{ij})
\underline{\varphi_{j}}(\omega)]
\label{2.8}
\end{eqnarray}
while the Kondo part of the action is decoupled
in the mean field approximation as in Ref. (\cite{15})
\begin{eqnarray}
A_{K}=\frac{\beta J_{K}}{N}\displaystyle\sum_{\sigma}
[\displaystyle\sum_{i\omega}\psi_{i\sigma}^{\dag}
(\omega)\varphi_{i\sigma}(\omega)][\displaystyle\sum_{i,\omega}
\varphi_{i-\sigma}^{\dag}\psi_{i-\sigma}(\omega)]
\label{2.9}
\end{eqnarray}
Here $ \beta$ is the inverse absolute temperature and we
also have
\begin{eqnarray}
A_{SG}=\displaystyle\sum_{i,j} J_{ij}S_{fi}^{z}S_{fj}^{z}
\label{2.10}
\end{eqnarray}
where in the static approximation \cite{15,16,17,24}
\begin{eqnarray}
  S^{z}_{fi}= \frac{1}{2}\displaystyle \sum_{\omega}
  \underline{\psi_{i}^{\dag}}(\omega)\underline{\sigma^{z}}
{\psi_{i}}(\omega)
\label{2.11}
\end{eqnarray}
The Kondo order is described by the complex order parameter
\begin{eqnarray}
\lambda_{s}^{\dag}=\frac{1}{N}\displaystyle
\sum_{i,\omega}\langle\psi_{is}^{\dag}(\omega)
\varphi_{is}(\omega)\rangle
\nonumber\\
\lambda_{s}=\frac{1}{N}\displaystyle
\sum_{i,\omega}\langle\varphi_{is}^{\dag}(\omega)
\psi_{is}(\omega)\rangle
\label{2.12}
\end{eqnarray}
that in a mean field theory \cite {6,15,26} describes the
correlations
$\lambda_{s}^{\dag}=\langle f_{is}^{\dag}d_{is}\rangle$
and  $\lambda_{s}=\langle d_{is}^{\dag}f_{is}\rangle$.
   The approximation used in Eq.(\ref{2.12}), which is equivalent to
the "slave boson" method, is certainly one of the best practical
methods used for the Kondo lattice problem.The Kondo temperature for the
lattice is determined here by the temperature at which $\lambda $
becomes equal to zero and this approximation gives a fairly reasonable
description of the Kondo phase. However, it is unable to give a "mixed"
 SG- Kondo phase, as it was previously shown in
the case of the antiferromagnetic-Kondo transition\cite{28}.
Except for the presence of the transverse field $\Gamma$ in $A_{0}$
in Eq. (\ref{2.8}), the calculation follows the same steps as in
Ref. (\cite{15}) and we refer the reader to this paper for details.
We show in Ref.(\cite{15})
 that standard manipulations give for the
averaged free energy within a replica symmetric theory:
\begin{eqnarray}
\beta F=2\beta J_{K}\lambda^{2}+\frac{1}{2}\beta^{2}J^{2}
(\overline{\chi}^{2}+2q\overline{\chi})
-\beta\Omega \label{2.13}
\end{eqnarray}
where
\begin{eqnarray}
\beta\Omega=\lim_{n \rightarrow 0}\frac{1}{Nn}\{\displaystyle
\int_{-\infty}^{+\infty}\displaystyle \prod_{j}^{N}Dz_{j}\displaystyle
 \prod_{\alpha}^{n}\displaystyle \int_{-\infty}^{+\infty}\displaystyle
\prod_{j}D\xi_{\alpha j}\exp{(\displaystyle
\sum_{\omega}\ln{|\underline{G}_{ij \alpha}^{-1}(\omega)|})}-1\}
\label{2.14}
\end{eqnarray}
and  $q, \overline{\chi}$, and $\lambda$ must be
taken at their saddle point value. Here $ \alpha $ is the
replica index and $q$ is the SG
order parameter\cite {15,18,24} while  $\chi=\beta \overline
{\chi}$ is the static uniform spin susceptibility
of the localized f-electrons.
\begin{eqnarray}
q_{\alpha\not=\beta}=q  =\lim_{n\rightarrow
0}\frac{1}{n(n-1)}\displaystyle\sum_{\alpha\not=\beta} \langle
S_{\alpha}^{z}S_{\beta}^{z}\rangle \, ,
\label{2.15}
\end{eqnarray}
\begin{eqnarray}
q_{\alpha\alpha}=q+\overline{\chi} =\lim_{n\rightarrow
0}\frac{1}{n}\displaystyle\sum_{\alpha} \langle
S_{\alpha}^{z}S_{\alpha}^{z}\rangle \, \,,
\label{2.16}       %
\end{eqnarray}
We use the notation $Dx=\frac{dx}{\sqrt{2\pi}}e^{-\frac{1}{2}x^2}$
and we refer the reader to Ref. (\cite{15}) for the details of
the calculation. The matrix $\underline{G}_{ij\alpha}(\omega)$
in Eq. (\ref {2.14}) is the time Fourier transform of the matrix
Green's function $\underline{G_{ij}}(\tau)=
 i\langle T \underline{\psi_{i}}(\tau)\underline{\psi_{j}^{\dag}}(0)
 \rangle$ for  the
 localized electrons in the presence of random fields  $z_{j}$ and
$\xi_{\alpha j}$ at every site, and  satisfies the
equation\cite{15}
\begin{eqnarray}
\underline{G_{ij\alpha}^{-1}}(\omega)=[i\omega-\beta\epsilon_{0}-
 \underline{\sigma_{z}} h_{j\alpha}+ \beta\Gamma\underline{\sigma_{x}}]
 \delta{ij} -\beta^{2}J_{K}^{2}\lambda^{\dagger}
 \lambda\gamma_{ij}(\omega)\underline{1}
\label{2.17}
\end{eqnarray}
where
\begin{equation}
h_{j\alpha}=\sqrt{2q}\beta Jz_{j}+\sqrt{2\overline{\chi}}\beta
J\xi_{\alpha j}
\label{2.18}
\end{equation}
while $ \gamma_{ij}(\omega)$ is the time Fourier transform of the
conduction electron Green's function
$\gamma_{ij}(\tau)=i\langle T\varphi_{is}(\tau)\varphi_{is}(0)
\rangle $
and is given by
\begin{equation}
\gamma_{ij}^{-1}=[i\omega-\beta\mu_{c}]\delta_{ij}-\beta t_{ij}
\label{2.19}
\end{equation}
We  obtained in Eq.\ (\ref {2.17})  the Green's function for the
f-electrons in a Kondo lattice\cite{15}, but now in the
presence of a random field $h_{j}$ at every site that prevents
us from diagonalizing
by means of a Fourier transformation. In the pure SG limit
$ J_{K}=0 $,
the Green's function in Eq. (\ref{2.17}) is local and the
integrals in
Eq. (\ref{2.14}) reduce to a one site problem, while in the
Kondo limit
$ J=0$ the random fields vanish and the integrals separate
in reciprocal
space, giving the known results for a Kondo lattice.
We adopt here a
decoupling approximation introduced in Ref. (\cite{15})
that reproduces
correctly these two limits, and corresponds to consider
independent Kondo
lattices, in place of the independent Kondo impurities
in Ref. (\cite{27}).
 We replace the Green's function
$\underline{G}_{ij\alpha}(\omega,h_{1\alpha}...h_{j\alpha}...
h_{N\alpha})$
by the Green's functions $\underline{G}_{\mu\nu}(\omega,h_{j\alpha})$
with, $j=1...N $, of $ N $ independent Kondo lattices,each one
with a "uniform"
field $h_{j\alpha}$ at every site $\mu, \nu$ by means of
the approximation
\begin{eqnarray}
\ln{|\underline{G}_{ij\alpha}^{-1}(\omega,h_{1}...h_{N})|}
\approx\frac{1}{N}\displaystyle\sum_{j}\ln{|\underline{G}_{\mu\nu}^
{-1}(\omega,h_{j\alpha})|}
\label{2.20}
\end{eqnarray}
where $\underline{G_{\mu\nu}}(\omega,h_{j\alpha})$ is the
f-electron
Green's function for a fictitious Kondo lattice that has
an uniform field
$h_{j\alpha}$ at every site $\mu,\nu$ and satisfies the equation:
\begin{eqnarray}
\underline{G}_{\mu\nu}^{-1}(\omega ,h_{j\alpha})=[(i\omega -
\beta\epsilon_{0})\underline{1}-
\underline{\sigma}_{z} h_{j\alpha}+\beta\Gamma\underline
{\sigma}_{x}]\delta_{\mu\nu}
-\beta^{2}J_{K}^{2}\lambda^{2}\gamma_{\mu\nu}\underline{1}
\label{2.21}
\end{eqnarray}
where from Eq.\ (\ref {2.19})
\begin{equation}
\gamma_{\mu\nu}(\omega)=\frac{1}{N}\displaystyle \sum_{k}\frac{1}
{i\omega-\beta\epsilon_{k}}e^{i\vec{k}\cdot\vec{R}_{\mu\nu}}
\label{2.22}
\end{equation}
Now Eq.\ (\ref {2.21}) may be easily solved by a Fourier transformation
with the result:
\begin{eqnarray}
\ln{|\underline{G_{\mu\nu}^{-1}}(\omega ,h_{j\alpha})|}=\displaystyle
\sum_{\vec{k}}\ln{|\underline{G_{\vec{k}}^{-1}}(\omega ,h_{j\alpha})|}
\label{2.23}
\end{eqnarray}
where
\begin{eqnarray}
\underline{G_{\vec{k}}^{-1}}(\omega ,h_{j\alpha})=
[i\omega-\beta\epsilon_{0}]\underline1-\underline{\sigma_{z}} h_{j\alpha}
+ \beta\Gamma\underline{\sigma_{x}}-
\beta^{2}J_{K}^{2}\lambda^{2}\frac{1}{i\omega-\beta\epsilon_{k}}.
\label{2.24}
\end{eqnarray}
We may now introduce Eq.\ (\ref {2.24}) and Eq.\ (\ref {2.20}) in
Eq.\ (\ref {2.14}), the integrals over the fields separate and we obtain
\begin{equation}
\beta\Omega=\displaystyle\int_{-\infty}^{+\infty}Dz\ln{\{\displaystyle
\int_{-\infty}^{+\infty}D\xi\exp{(\displaystyle
\frac{1}{N}\displaystyle\sum_{\vec{k}}\sum_{s}S_{s}(\vec{k},H))}}\}
\label{2.25}
\end{equation}
with
\begin{equation}
\sum_{s}S_{s}(\vec{k},H)=\displaystyle\sum_{\omega}
\ln{[\underline{G_{\vec{k}}}^{-1}(\omega,h)]}
\label{2.26}
\end{equation}
and $h$ is  given in Eq.\ (\ref {2.18}), with $z$ and $\xi$
in place of $z_{j}$ and $\xi_{j\alpha}$, while
\begin{equation}
H=\sqrt{h^2+ (\beta\Gamma)^2}
\label{2.27}
\end{equation}
The sum over the fermion frequencies is performed in the standard way
by integrating  in the complex plane\cite{25}, with the result:
\begin{equation}
S_{s}(\vec{k},H)=\ln{[(1+e^{-\omega_{s^{+}}})
(1+e^{-\omega_{s^{-}}})]}
\label{2.28}
\end{equation}
where
\begin{equation}
\omega_{s^{\pm}}=\frac{1}{2}[\beta\epsilon_{k}+s H]\pm
\{\frac{1}{4}(\beta\epsilon_{k}-s H)^{2}+
(\beta J_{K}\lambda)^{2}\}^{\frac{1}{2}}.
\label{2.29}
\end{equation}
We consider $\epsilon_{0}=0$ that corresponds to an average
occupation $\langle n_{f}\rangle=1$, per site.
Replacing sums by integrals , in the approximation of a constant
density of states for the conduction band electrons, $ \rho(\epsilon)=
\rho=\frac{1}{2D}$ for $-D<\epsilon <D$, we obtain from
Eq.\ (\ref {2.26}) to Eq.\ (\ref {2.29}) the final expression for the
free energy in Eq.\ (\ref {2.13}):
\begin{eqnarray}
\beta F=2\beta J_{K}\lambda^{2}+\frac{1}{2}\beta^{2}J^{2}
(\overline{\chi}^{2}+2q\overline{\chi})-
\displaystyle\int_{-\infty}^{+\infty}Dz\ln{\{\displaystyle
\int_{-\infty}^{+\infty}D\xi e^{E(H)}\}}
\label{2.30}
\end{eqnarray}
with
\begin{eqnarray}
E(H)=\frac{1}{\beta D}\displaystyle\int_{-\beta D}^{+\beta D}dx
\ln{\{\cosh{\frac{(x+H)}{2}}+\cosh{(\sqrt{\Delta})}\}},
\label{2.31}
\end{eqnarray}
\begin{eqnarray}
\Delta=\frac{1}{4}(x-H)^2+(\beta J_{K}\lambda)^2
\label{2.32}
\end{eqnarray}
and from Eq.\ (\ref {2.27}) we have
\begin{eqnarray}
H=\sqrt {(\beta\Gamma)^2+
(\beta J)^2(\sqrt{2q}z+\sqrt{2\overline{\chi}}\xi)^{2}}
\label{2.33}
\end{eqnarray}
\section*{3.\ Results}
 The SG-Kondo transition is described in the present model by
the three parameters $ J_{K},J $  and $ \Gamma$, which cannot be
considered here as completely independent from each other.
In fact, the Kondo effect and the RKKY interaction originate
from the same intra-site exchange interaction, but the
necessity of considering an additional inter-site exchange term
has been already recognized\cite{6,28} and a relationship
giving the intersite exchange parameter varying as $ (J_{K})^2$
has been introduced in ref.\cite{6}. Since the transverse
field is introduced here to mimic the spin flipping part
of the Heisenberg Hamiltonian that originate in the RKKY
interaction, we assume a similar relationship
$ \Gamma= \alpha (J_{K})^2 , \alpha \le 1$ to have a better
description of the SG-Kondo transition. In particular, the
consideration of the previous relationship for $ \Gamma $
will be important for the existence of the QCP and the
comparison with experiment.
The saddle point equations for $\overline{\chi}, q $ and $\lambda$
are obtained
by extremizing $ \beta F$ with respect to these order parameters.
By making $ q=\lambda=0$ in the saddle point equations we obtain the
equations for the second order critical lines $T_{1}(J_{K})$ and
$T_{2}(J_{K})$ that separate the paramagnetic phase from the spin-glass
and Kondo phases, respectively. They are given by:
\begin{eqnarray}
\overline{\chi}(\beta)=\displaystyle \frac{1}{1+J_{0}}\int_{-\infty}^{+\infty}D\xi
\xi^{2}\sinh(H_{0})/H_{0}
\label{2.34}
\end{eqnarray}
where
\begin{eqnarray}
 J_{0}=\displaystyle \int_{-\infty}^{+\infty}D\xi \cosh(H_{0})
\label{2.35}
\end{eqnarray}
and $H_{0}= H(q=0)$.\\
Thus we obtain for the spin glass transition temperature
$T_{1}(J_{K})$\cite{24}:
\begin{eqnarray}
\overline{\chi}(\beta_{1})= \frac{1}{\sqrt{2}\beta_{1}J}
\label{2.36}
\end{eqnarray}
that together with Eq.\ (\ref{2.34}) gives $T_{1}$ and
$ \overline{\chi}(\beta_{1})$, while the Kondo transition
temperature $T_{2}(J_{K})$ is
given by
\begin{eqnarray}
1-\frac{\beta_{2} J_{k}}{4(1+J_{0})}\displaystyle
\int_{-\infty}^{+\infty} D\xi \cosh(H_{0}/2) \nonumber\\
\times\frac{1}{\beta_{2} D}\displaystyle\int_{-\beta_{2}D}^{+\beta_{2}D}
dx\frac{1}
{\cosh(x/2)}
\frac{\sinh(\frac{H_{0}-x}{2})}{\frac{H_{0}-x}{2}}=0
\label{2.37}
\end{eqnarray}
that must be solved together with Eq.\ (\ref {2.34}).\\
The numerical solution of the saddle point equations as a
function of $ \frac{T}{J }$ and $ \frac{J_{K}}{J} $ yields
the phase diagram in Fig.1, where we considered $D/J = 12$
and two values of $\alpha J $.  The two cases corresponding to
these two values of $\alpha J $ are given in Fig.1 .
The first case (represented by full lines) corresponds to
the really interesting situation of the phase diagram with two QCP,
while the second case (represented by the dashed line $ T'$)  is
plotted here for comparison to show the case without QCP.\\
 In the first case, corresponding to
$ \alpha J = 0.01348 $ (solid line) the second order critical line
$ T_{1}(J_{K})$ that separates the paramagnetic
from the spin glass phase ends at a QCP corresponding to
a $ J_{K}$ value called $J_{K1}^{c}$ here, while the
second order critical line $T_{2}(J_{K})$ that separates the
paramagnetic from the Kondo phase ends at a second QCP
$J_{K2}^{c}$ and we have here $J_{K2}^{c} >J_{K1}^{c}$.\\
  An asymptotic calculation at low temperatures gives:
\begin{eqnarray}
k_{B}T_{1}= \frac{\sqrt{2}}{3}[2\sqrt{2}J-\alpha J_{K}^{2}]
\label{2.38}
\end{eqnarray}
and the corresponding value for $J_{K1}^{c}$ at
the QCP is given by $\alpha (J_{K1}^{c})^{2}=2\sqrt{2}J$.\\
On the other hand, close to the second QCP an asymptotic
calculation gives:
\begin{eqnarray}
k_{B}T_{2} \approx \log{(T_{K}^{-1}\frac{\alpha(J_{K})^2}{1+\alpha(J_{K})^2/D})}
\label{2.39}
\end{eqnarray}
where $T_{K}= D\exp{(-2D/J_{K})}$. Thus the value of $J_{K2}^{c}$
is the solution of $T_{K}(J_{K2}^{c}) = \alpha(J_{K2}^{c})^2 $
in the asymptotic limit of very large D values.
 In order to obtain a solution $ J_{K2}^{c} $
that makes $ T_{2}=0 $ in Eq.\ (\ref{2.39}) it is necessary that
$ 4\alpha D = 0.6473 $,
what gives $ \alpha J = 0.013485 $ for  $D= 12J $, and for
these values of the parameters
in Fig.1, we have obtained two QCP with $J_{K2}^{c} > J_{K1}^{c}$.
 In the second case corresponding to a smaller value of
$ \alpha , \alpha J = 0.1344$, the two QCP disappear and we
obtain a single transition line ( dashed line $ T'$) separating the
paramagnetic from the ordered ( spin-glass and Kondo) phases.
The physical case shown
in Fig.1, where the two QCP are very close to each other
and almost equal, corresponds well to experimental phase
diagrams obtained in some Cerium or Uranium disordered alloys.
We obtain from Eq.\ (\ref {2.34})
in the disordered region that
 $ \overline{\chi} \rightarrow 0.5$ when $ \beta  \rightarrow 0$,
 thus giving the expected Curie-Weiss behaviour for the static
susceptibility $ \chi = \beta \overline{\chi}$,
while for  $ \beta  \rightarrow  \infty $ we get
$ \chi \rightarrow  \Gamma + \sqrt{\Gamma^{2} -8J^{2}}$
when $\Gamma = \alpha J_{K}^{2}  > 2\sqrt{2} J$.
 We recover here the square root singularity in the linear
static susceptibility that is present in other quantum spin
glass models\cite{20,21,22}.
It is interesting to remark that $ \overline{\chi} =\chi T= S(S+1)$
is the residual effective localized spin in Curie-Weiss theory.
The value for $\overline{\chi} $ on the critical line $T_{2}(J_{K})$
is presented in Fig.1, where we notice the steep drop to almost zero
close to the QCP $ J_{K2}^{c} $.

\section*{4.\ Conclusions}
We study in this paper the phase transitions in a heavy fermion system
represented by a Hamiltonian that couples the $S_{z}$
components of the localized
spins of a Kondo lattice \cite{7}with random, long range
interactions,like in the SK model for a spin glass\cite{18},
while the transverse components are acted upon by
a field in the x-direction\cite{24}. As the transverse field $\Gamma$
mimics the spin flip part of the Heisenberg coupling among
localized spins, that originates in the RKKY interaction,
we assume that $ \Gamma \approx \alpha J_{K}^2 $ where $J_{K}$
is the antiferromagnetic Kondo coupling\cite{6}.
Using functional integrals techniques and a static, replica
symmetric Ansatz for the Kondo and spin glass order parameters,
we derive a mean field expression for the free energy and the
saddle point equations for the order parameters.
The use of the static Ansatz  in the
case of the transverse spin glass is justified,
 because it is the singularity of the zero frequency mode that
determines the critical line\cite{20,21,22}.
 Numerical solution of the saddle-point equations allows us to draw the magnetic
phase diagram in the $J_{K} vs T$ plane( presented in Fig.1)
for fixed values
of $J$ and $D$ with $ D/J = 12 $ and for two values of $ \alpha J ,
\alpha J = 0.01348 $   or $ \alpha J = 0.01344. $ \\
  Figure 1 shows three distinct phases. At high temperatures, the
"normal" phase is paramagnetic with vanishing Kondo and spin glass
order parameters, i.e. $\lambda= q= 0$. When temperature is lowered,
for not too large values of the ratio $J_{K}/J$, a second-order transition
 line is found at $T= T_{1}$ to a spin glass phase with $q >0$ and
$\lambda= 0$.The critical line $T_{1}(J_{K})$ ends at a quantum critical
point  $J_{K1}^{c}$.Finally, for large values of the ratio
$J_{K}/J \ge J_{K2}^{c}/J$,
where $J_{K2}^{c} \ge J_{K1}^{c}$, there is a second order transition line
to a Kondo state with $\lambda \ge 0$.
These results are very sensitive to the value
of $D/J$ and $\alpha J$. For $D/J = 12 $  and
$\alpha J \le 0.01344 $
 the QCP disappear
while they are favored for
 larger values of $\alpha J$.
 We can also remark that we get here
only "pure"
Kondo or SG phases and never a mixed SG-Kondo phase with the
two order parameters
different from zero  as already noticed for the competition
between Kondo and antiferromagnetic phases\cite{28};
this result is probably connected to the
approximations used
here to treat the starting Hamiltonian.
A QCP has been observed experimentally in several Cerium and Uranium
disordered alloys\cite{12,13,14} and the phase diagram shown in Fig.1 which
yields two QCP improves considerably the description of the spin
glass-Kondo transition with respect to previous publications\cite{15,16,17}.
However, the experimental situation is generally not really clear;
for example, there is no experimental information on the precise
nature of the SG-Kondo transition in $CeNi_{1-x}Cu_{x}$ alloys\cite{10}.
Moreover the phase diagrams of several systems involve also an
antiferromagnetic phase and this case is theoretically studied
elsewhere\cite{17}. Our theoretical results describe also the spin glass
and Kondo phases in Uranium alloys.
 An unsolved basic question concerns also the existence or not of
 a "mixed"
SG-Kondo phase in Cerium  and Uranium
disordered alloys and this problem is worth of being studied
experimentally in more detail.
Thus, further experimental work is necessary, but our model yields a new
striking point in the behaviour of heavy fermion disordered alloys in the
vicinity of the quantum critical point.
\section*{Acknowledgement}
We acknowledge enlightening discussions with J. C. G\'omez Sal.
 This work is partially supported by
Conselho Nacional de Pesquisa e Desenvolvimento (CNPQ) and by
Financiadora de Estudos e Projetos (FINEP).One of us (B.C.) would
like to thank the French-Brazilian CNRS-CNPq  cooperation.
\newpage

\newpage
\section*{Figure captions}
Figure 1:Phase diagram in the $T- J_{K}$ plane
as a function of $T/J$ and $J_{K}/J$ for the transverse field
$\Gamma= \alpha J_{K}^2, \alpha J = 0.01348 $ ( solid line) and
$\alpha J= 0.01344 $ (dashed line),
for the conduction bandwidth $ D/J =12$. The critical (solid)
second order
line $ T_{1} $ for low values of $J_{K}$ separates the paramagnetic
phase ($q=\lambda=0$) for
high temperatures from the spin-glass phase ($q \ge 0, \lambda=0$) at
low temperatures and ends
at a QCP. The critical second order (solid) line
$ T_{2}$ for large
values of $J_{K}$ separates the paramagnetic phase from the Kondo phase
$(\lambda \ge 0, q=0)$ and ends at a second QCP.
 The critical ( dashed) line T'does not have a QCP.
 The dash-dotted line represents
the "pure" Kondo temperature $T_{K}$ and the dotted
line represents the residual magnetic moment $ \overline{\chi}_{c}(J_{K}) $
on the line $ T= T_{2} $.
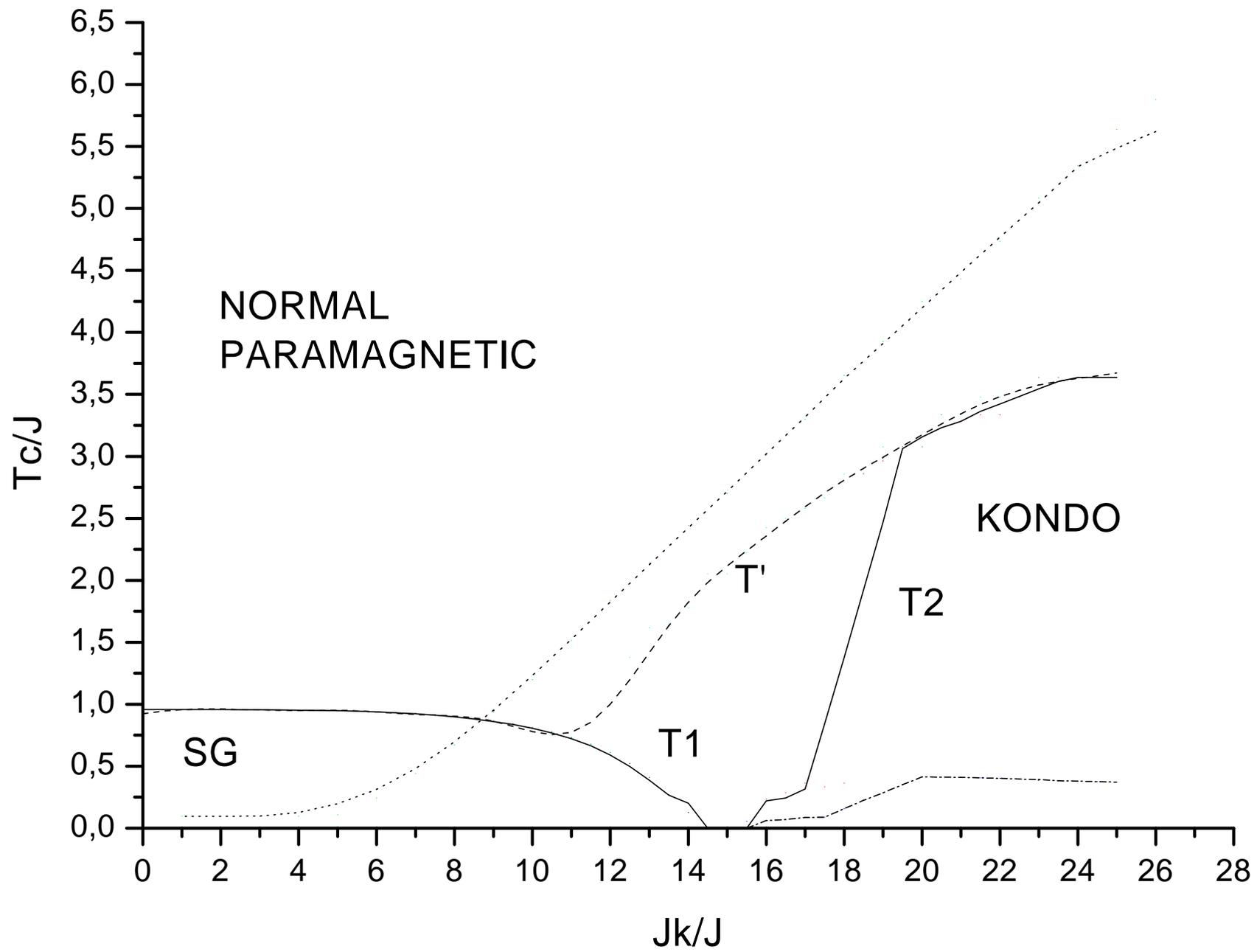
\end{document}